\documentclass{jaa}
\usepackage{txfonts}
\usepackage{graphicx}
\usepackage{psfig}

\usepackage[authoryear]{natbib}
\begin{document}
%                 So begins the second great struggle...                %

\title[Spectral variability in RLQs] %% give here short title %%
{The spectral variability in radio-loud quasars}\author[Minfeng Gu]%
       {Minfeng Gu$^*$\\
%       \thanks{e-mail: gumf@shao.ac.cn}\\
         Key Laboratory for Research in Galaxies and Cosmology, Shanghai Astronomical\\
         Observatory, Chinese Academy of Sciences, 80 Nandan Road Shanghai 200030, China\\
$^*$e-mail: gumf@shao.ac.cn}
   \maketitle
\label{firstpage}

\begin{abstract}
The spectral variability of a sample of 44 flat-spectrum radio quasars (FSRQs) and 18 steep-spectrum radio quasars (SSRQs) in the SDSS stripe 82 region is investigated. Twenty-five of 44 FSRQs show a bluer-when-brighter trend (BWB), while only one FSRQ shows a redder-when-brighter trend, which is in contrast to our previous results. Eight of 18 SSRQs display a BWB. We found an anti-correlation between the Eddington ratio and the variability amplitude in r band for SSRQs, which is similar to that in radio-quiet AGNs. This implies that the thermal emission from the accretion disk may be responsible for the variability in SSRQs. The spectral variability from the SDSS multi-epoch spectroscopy also shows BWB for several SSRQs, consistent with that from photometry.
%Moreover, we present the optical continuum variability of radio-loud broad absorption-line quasars (BALQs) for the first time %for two sources with steep radio spectra.
\end{abstract}

\begin{keywords}
galaxies: active --- quasars: general --- galaxies: photometry --- galaxies: spectroscopy
\end{keywords}

\section{Introduction}

Active galactic nuclei (AGNs) exhibit variability at almost all
wavelengths. The radido-loud quasars are 
divided into two populations, flat-spectrum radio quasars (FSRQs) and 
steep-spectrum radio quasars (SSRQs). In FSRQs, the non-thermal emission
 from a relativistic jet usually are dominant and Doppler boosted, due to the small viewing angle. 
In contrast, the SSRQs are usually lobe-dominated
radio quasars, with a large viewing angle, therefore the
 beaming effect is not severe. 

Although a bluer-when-brighter 
trend is commonly observed in blazars (e.g. Fan et al. 1998; Raiteri et al. 2001; Villata et al. 2002; Wu
et al. 2007), the opposite trend of redder-when-brighter 
has also been found (e.g. Gu et al. 2006; Dai et al. 2009; Rani et al. 2010; Bian et al. 2012), especially in FSRQs (e.g. Gu et al. 2006). However, it is unclear
whether a redder-when-brighter trend is generally present in FSRQs. Moreover, 
the optical and color/spectral variations
of SSRQs have been poorly studied, and the variability mechanism is largely unknown. For these reasons, 
we investigate the optical variability and the 
spectral variability for a sample of radio-loud quasars (see details in Gu \& Ai 2011a,b). 

\begin{figure}[b]
% \vspace*{-2.0 cm}
\begin{center}
 \includegraphics[width=0.7\textwidth]{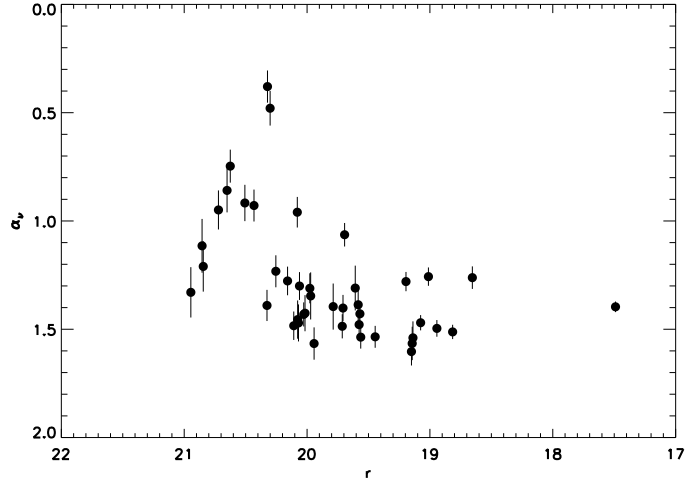}
% \vspace*{-1.0 cm}
 \caption{The relationship between the spectral index and the PSF magnitude
   at r band for SDSS J001130.40$+$005751.7 ($z=1.4934$). A significant anti-correlation is
   present, which implies a redder-when-brighter trend.}
   \label{fig1}
\end{center}
\end{figure}

\section{Sample}

Our sample of 62 radio-loud quasars consists of 44 FSRQs and 18 SSRQs. The initial quasar
sample was selected as those quasars in both the SDSS DR7 quasar
catalogue (Schneider et al. 2010) and Stripe 82 region. The Stripe-82 region, i.e. right ascension $\alpha = 20^{\rm h} -
4^{\rm h}$ and declination $\delta=-1^{\circ}.25 - +1^{\circ}.25$,
was repeatedly scanned during the SDSS-I phase (2000 - 2005) and
also over the course of three 3-month campaigns in three successive
years in 2005 - 2007 known as the SDSS Supernova Survey. We
cross-correlate the initial quasar sample with the Faint Images of
the Radio Sky at Twenty centimeters (FIRST) 1.4-GHz radio catalogue (Becker, White \& Helfand 1995), the Green Bank 6-cm (GB6) survey at
4.85 GHz radio catalogue (Gregory et al. 1996), and the
Parkes-MIT-NRAO (PMN) radio continuum survey at 4.85 GHz (Griffith \& Wright, 1993). The radio spectral
index $\alpha_{\nu}$ was then calculated between the single or
integrated FIRST and/or NRAO VLA Sky Survey (NVSS) 1.4 GHz and either or both of the GB6 and PMN 4.85 GHz. 
We define a quasar to be a SSRQ according to
its radio spectral index $\alpha_{\nu}>0.5$ ($f_{\nu} \propto \nu^{-\alpha_{\nu}}$), and otherwise as 
FSRQs.

\section{Results}

\begin{figure}[b]
% \vspace*{-2.0 cm}
\begin{center}
 \includegraphics[width=0.7\textwidth]{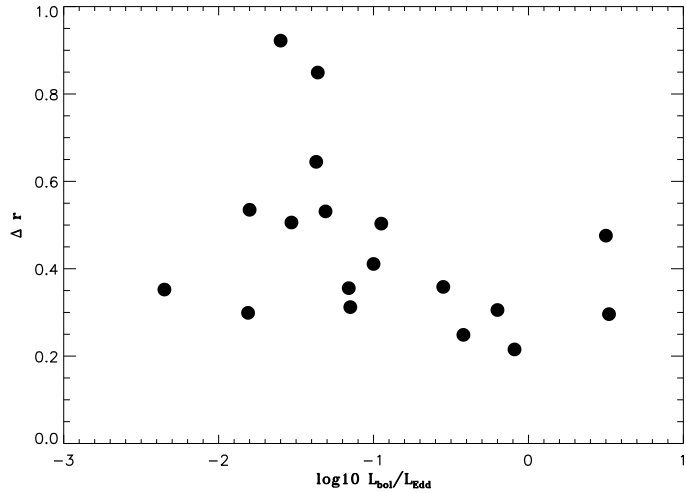}
% \vspace*{-1.0 cm}
 \caption{The Eddington ratio versus the variability at r band $\Delta r$ for 
SSRQs.}
   \label{fig2}
\end{center}
\end{figure}

\subsection{The spectra variability from photometry}
We directly used the point-spread-function magnitudes in the CAS Stripe82 database from
the photometric data obtained during the SDSS-I phase from data
release 7 and the SN survey during 2005
- 2007. We found that radio-loud quasars all show more or less variations, ranging
from 0.18 to 3.46 mag at r band. FSRQs show more pronounced variations than SSRQs, with 
$\Delta \rm  r>1.0$ mag in four FSRQs while none in SSRQs. By performing the correlation 
analysis between the spectral index and r band brightness, the redder-when-brighter trend is
only found in one FSRQ SDSS J001130.400+005751.8 ($z=1.4934$) (see Fig. \ref{fig1}), which could be explained by the 
thermal accretion disk emission as other FSRQs (Gu et al. 2006). In contrast, the bluer-when-brighter trend is more common 
in FSRQs (25 out of 44 sources), and in SSRQs (8 of 18) as well. The results of FSRQs are in contrast
to our previous results that FSRQs generally show the redder-when-brighter trend (Gu et al. 2006, see also Rani et al. 2010). For all SSRQs studied, we found an
anti-correlation between the Eddington ratio and the variability
amplitude in r band (Fig. 2), which is similar to that in radio-quiet AGNs.
This implies that the thermal emission from the accretion disk may
be responsible for the variability in SSRQs. 
%We augue that a disk-like BLR geometry may present
% in our sample, with a broader Mg II at steeper radio spectral index.

\subsection{The spectral variability from spectroscopy}

\begin{figure}[b]
% \vspace*{-2.0 cm}
\begin{center}
 \includegraphics[width=0.7\textwidth]{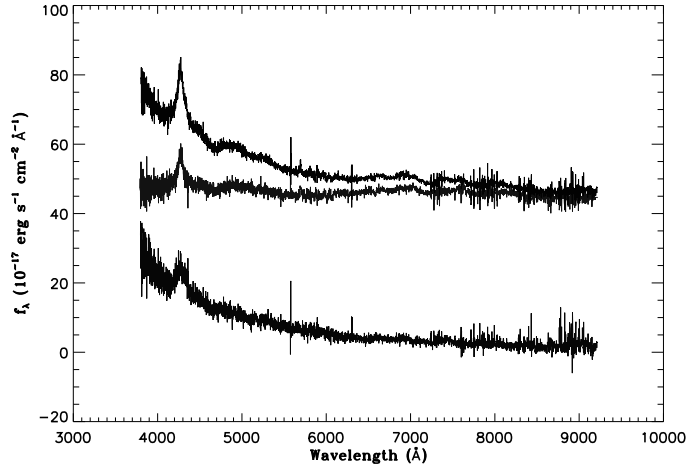}
% \vspace*{-1.0 cm}
 \caption{The SDSS spectra of SDSS J022508.07+001707.2 ($z=0.527$) in the observed frame. The up two shows the spectra at two epochs shifted up by a flux value 40.0 to separate from the bottom one, which is the difference 
between up two showing the variabibility between two epochs.}
   \label{fig3}
\end{center}
\end{figure}

In 18 SSRQs, we have collected the multi-epoch spectroscopy for nine sources from
SDSS.  The significant variations were found in five sources. All these five sources show BWB, 
which is consistent with the results from photometry (Gu \& Ai 2011b). The BWB has also been 
commonly found in a large sample of quasars selected from SDSS with multi-epoch spectroscopy 
(Guo \& Gu 2013). However, the RWB was found in a large fraction of FIRST bright quasars (Bian et al. 2012), 
even in radio quiet quasars.

For our sample, an example is shown in Fig. 3 for
SDSS J022508.07+001707.2 ($z=0.527$). It can be directly seen from the spectra that the 
source becomes bluer when it goes brighter, which is also evident from the spectrum 
difference. Intriguingly, the residual of Mg II line is apparent in the spectrum difference. 
As a matter of fact, the narrow lines are less variable, since the narrow line region locates typically at kpc scales, far from the nuclei. Therefore, the broad profile of the residual Mg II line implies that the broad Mg II line is also varying along with the variations in the continuum. This is in good agreement with the photoionization model and that the thermal emission 
is responsible for the continuum, although there might be certain time delay between their variations. \\

\noindent{{\textbf{Acknowledgements}}} \\
\noindent{This work is supported by the 973 Program (No. 2009CB824800), and by the NSFC grant 11073039. }

\label{lastpage}
\end{document}